\documentclass[aps,prc,onecolumn,superscriptaddress,showpacs,nofootinbib,floatfix]{revtex4}
\usepackage[dvips]{graphics,graphicx}
\usepackage{amsmath,amssymb}
\begin{document}
\preprint{WU-HEP-03-6}
\title{Source Chaoticity from Two- and Three-Pion Correlations in Au+Au
collisions at $\mathbf{\sqrt{s_{NN}}=130}$ GeV}
\author{Kenji Morita}
\email{morita@hep.phys.waseda.ac.jp}
\affiliation{Department of Physics, Waseda University, Tokyo 169-8555,
Japan}
\author{Shin Muroya}
\email{muroya@yukawa.kyoto-u.ac.jp}
\affiliation{Tokuyama University, Shunan, Yamaguchi 745-8511,
Japan}
\author{Hiroki Nakamura}
\email{naka@hep.phys.waseda.ac.jp}
\affiliation{Department of Physics, Waseda University, Tokyo 169-8555,
Japan}
\date{\today}
\begin{abstract}
 We consistently analyze two-pion correlation functions and 
 three-pion correlation functions of 130$A$ GeV Au+Au collisions
 measured at RHIC, applying three models of partially coherent pion sources.
The effect of long-lived resonances on the chaoticity
 in the
 two-pion correlation function is estimated on the basis of a statistical
 model. We find that the chaoticity extracted from the three-pion
 correlation function is consistent with that extracted from the two-pion
 correlation function at vanishing relative momenta after the subtraction
 of the contribution of the long-lived resonance decays.
 All three models indicate that pions are emitted from the mixture of
 a chaotic source and coherent sources in the 130$A$ GeV Au+Au
 collisions.
\end{abstract}

\maketitle

\section{Introduction}\label{sec:intro}
Pion interferometry is one of the most powerful tools in relativistic
heavy ion physics. The fact that two-boson intensity correlation
functions provide information concerning the geometry of particle sources
is known as the Hanbury Brown-Twiss (HBT) effect. The
two-pion correlation function
$C_2(\boldsymbol{p_1,p_2})$ has been analyzed extensively to explore
space-time structures of collision dynamics
\cite{Tomasik_qgp3}. The HBT effect is based on the assumption
of a chaotic source; i.e., particles are emitted with
relatively random phases. The chaoticity index $\lambda$ in the two-particle
correlation function is defined as the
two-pion correlation strength, $\lambda = C_2(\boldsymbol{p,p})-1$. It is
unity for a chaotic source and vanishes for a coherent one. 
Thermal emission is naturally regarded as a perfect chaotic source,
\cite{Hama_PRD37,Morita_PRC61} but non-thermal components need not be
perfectly chaotic. For example, if a disoriented chiral condensation
(DCC) region is produced,
this domain can decay into coherent pions.
Therefore, the chaoticity index $\lambda$ is an important parameter
for understanding multiple particle production.
However, experimental data are not so simply understood; even for a
chaotic source, the chaoticity can be reduced by some other effects, such
as long-lived resonances
\cite{Gyulassy_PLB217,Bolz_PRD47,Csorgo_ZPHYS,Heiselberg_PLB,Wiedemann_PRC56}.

As an alternative tool to investigate particle sources,
the three-particle correlation function has also been proposed
\cite{Biyajima_PTP84,Heinz_PRC56,Nakamura_PRC60,Nakamura_PRC61}.
Regarding the chaoticity of the sources, the three-particle correlation is
a more promising tool than the two-particle correlation, because
long-lived resonances do not affect a normalized three-pion correlator
(third order cumulant correlation function),

\begin{align}
 r_3(\boldsymbol{p}_1,\boldsymbol{p}_2,\boldsymbol{p}_3)
=\frac{[C_3(\boldsymbol{p}_1,\boldsymbol{p}_2,\boldsymbol{p}_3)-1]
-[C_2(\boldsymbol{p}_1,\boldsymbol{p}_2)-1]
  -[C_2(\boldsymbol{p}_2,\boldsymbol{p}_3)-1]
-[C_2(\boldsymbol{p}_3,\boldsymbol{p}_1)-1]}
  {\sqrt{[C_2(\boldsymbol{p}_1,\boldsymbol{p}_2)-1]
[C_2(\boldsymbol{p}_2,\boldsymbol{p}_3)-1]
[C_2(\boldsymbol{p}_3,\boldsymbol{p}_1)-1]}}. 
 \label{eq:r3-1}
\end{align}
Here, $C_2(\boldsymbol{p_1},\boldsymbol{p_2})$ and
$C_3(\boldsymbol{p_1},\boldsymbol{p_2},\boldsymbol{p_3})$ are the two-
and three-pion correlation functions defined as
\begin{equation}
 C_2(\boldsymbol{p_1},\boldsymbol{p_2})
  =
  \frac{W_2(\boldsymbol{p_1},\boldsymbol{p_2})}{W_1(\boldsymbol{p_1})W_1(\boldsymbol{p_2})},
\end{equation}
and
\begin{equation}
 C_3(\boldsymbol{p_1},\boldsymbol{p_2},\boldsymbol{p_3})
 =
 \frac{W_3(\boldsymbol{p_1},\boldsymbol{p_2},\boldsymbol{p_3})}{W_1(\boldsymbol{p_1})W_1(\boldsymbol{p_2})W_1(\boldsymbol{p_3})},
\end{equation}
with $W_n(\boldsymbol{p_1},\cdots,\boldsymbol{p_n})$ being the $n$ particle
distribution.
The chaoticity index in the three-pion correlator is the weight
factor $\omega = r_3(\boldsymbol{p},\boldsymbol{p},\boldsymbol{p})/2$,
which becomes unity for a chaotic source.
The three-pion correlator used in the current experiments is a modified
form of Eq.~\eqref{eq:r3-1}, introduced due to a lack of statistics, as
\begin{equation}
 r_3(Q_3)=\frac{[C_3(Q_3)-1]-[C_2(Q_{12})-1]-[C_2(Q_{23})-1]-[C_2(Q_{31})-1]}
  {\sqrt{[C_2(Q_{12})-1][C_2(Q_{23})-1][C_2(Q_{31})-1]}}, \label{eq:r3}
\end{equation}
where $Q_{ij}=\sqrt{-(p_i-p_j)^2}$ and $Q_3=\sqrt{Q_{12}^2+Q_{23}^2+Q_{31}^2}$,
and $\omega$ is defined as $r_3(0)/2$. In both definitions, $\omega$ is
expected to approach the same value at small $Q_3$.

In this paper, we analyze both the three- and two-pion
correlation data measured by the STAR collaboration \cite{STAR_3pi} for
central events in Au+Au collisions at 130$A$ GeV using three kinds of
partially coherent models. The STAR
collaboration, adopting a partial coherent model for the pion source,
concludes that about 80\% of pions are emitted from a chaotic source
in the central collision events, and the chaoticity depends on the centrality
\cite{STAR_3pi}. In Ref.~\cite{STAR_3pi}, analysis is carried out only
with the
three-pion correlation function, because the two-pion correlation function
suffers from the effect of long-lived resonance decays. Hence, the
chaoticity extracted
from the three-pion correlation differs from that measured
as the strength of the two-pion
correlation function. One of the aims of this work is to account for this
discrepancy. After an appropriate correction for the long-lived resonance
decay contribution to the two-pion correlation function, the ``true''
chaoticity of the two-pion correlation function thereby obtained should
be consistent
with that extracted from the
three-particle correlation function. We show that the chaotic
fraction and the number of coherent sources evaluated from the ``true''
chaoticity of the two-pion correlation function are consistent with those
obtained from the three-pion correlation function. Then, we study the
structure of the source using three kinds of partial coherent source
models.
This paper is organized as follows. We give a brief explanation of the
three kinds of source models in the next section. In
Sec.~\ref{sec:lambda}, we make corrections to the two-pion correlation for
the long-lived resonance decay contributions. Section \ref{sec:omega} is
devoted to a combined analysis of two- and three-pion correlation
functions and analyses using the three models.

\begin{figure}[ht]
 \begin{center}
 \includegraphics{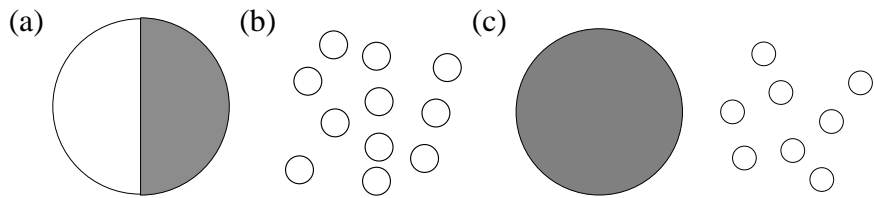}
 \end{center}
 \caption{\label{fig:models}Schematic depiction of Model I-III. The
 shaded areas represent the chaotic sources. The open area and
 small circles represent the coherent sources. (a) Model I, (b) Model
 II, (c) Model III.}
\end{figure}

\section{Models of pion emission}\label{sec:model}
Let us start by giving a brief explanation of the models. In Model I,
a chaotic source and a coherent source are assumed to be produced in
every event of the collisions. This is a very limited case, because, in
general, more
than one coherent source can be produced in a collision. In Model II, 
we consider the case in which multiple coherent sources are produced. Model III
is a mixture of Models I and II. A schematic depiction of the models is
given in Fig.~\ref{fig:models}.

Model I is a partially coherent model \cite{Heinz_PRC56} in which the
solitary parameter
$\varepsilon_{\text{pc}}$ represents the ratio of the particle number emitted
from the chaotic source to that emitted from both sources. In this
model, pions are
emitted from a mixture of a chaotic (thermal) and a coherent
source. Here we do not discuss the origin of the coherent component. 
In general, we need to fix
the source function of the coherent component, as well as that of the chaotic
one, in order to express the correlation functions. 
Fortunately, because the chaoticity $\lambda$ and the weight factor
$\omega$ are given by the correlation functions at vanishing relative
momenta, they can be expressed in terms of the chaotic fraction as
\begin{equation}
 \lambda = \varepsilon_{\text{pc}}(2-\varepsilon_{\text{pc}}), 
\quad
 \omega = \sqrt{\varepsilon_{\text{pc}}}
 \frac{3-2\varepsilon_{\text{pc}}}{(2-\varepsilon_{\text{pc}})^{3/2}}.
 \label{eq:ome-pc}
\end{equation}
Model II is a multicoherent source model,
\cite{Nakamura_PRC61} in which
the source is composed of a large number of similar, independent
coherent sources. Here we do not discuss the origin of the
coherence. Because each small coherent source is independent, a chaotic
source is realized as a cluster of an infinite number of the coherent sources.
Here, $\lambda$ and $\omega$ are given by
\begin{equation}
 \lambda = \frac{\alpha_{\text{m}}}{\alpha_{\text{m}}+1},
  \quad
 \omega = \frac{1}{2}\frac{2\alpha_{\text{m}}^2+2\alpha_{\text{m}}+3}
 {\alpha_{\text{m}}^2+3\alpha_{\text{m}}+1}
 \sqrt{\frac{\alpha_{\text{m}}+1}{\alpha_{\text{m}}}} \label{ome-m},
\end{equation}
where $\alpha_{\text{m}}$ represents the mean number of coherent
sources, which obeys the Poisson distribution.
Note that each of the above two models contains only a single parameter
($\varepsilon_{\text{pc}}$ for Model I and $\alpha_{\text{m}}$ for 
Model II) for two quantities, 
the chaoticity $\lambda$ and the weight factor $\omega$.
Model III is a mixture of these two
\cite{Nakamura_PRC61} [see also Fig.~\ref{fig:models}(c)] and has
already been applied to the SPS data by one of the present authors
(H.~N.). \cite{Nakamura_PRC66} This model contains two
parameters, which are related to $\lambda$ and $\omega$ as
\begin{align}
 \lambda &= \frac{\alpha}{\alpha+(1-\varepsilon)^2},\label{l-pm} \\
 \omega &= \frac{2\alpha^2 + 2\alpha 
 (1-\varepsilon)^2 + 3(1-\varepsilon)^3 (1-2\varepsilon)}
 {2[\alpha^2  + 3\alpha (1-\varepsilon)^2  +(1-\varepsilon)^3]} 
 \sqrt{\frac{\alpha+(1-\varepsilon)^2}
 {\alpha}},\label{ome-pm}
\end{align}
where $\varepsilon$ and $\alpha$ are the ratio
of the number of particles from the chaotic source to the total number
and the mean number of coherent sources, respectively.

Substituting $\lambda$ and $\omega$ from the experimental data 
into the above equations, we
can obtain the parameters of the models, $\varepsilon$ and
$\alpha$. In Models I and II, $\lambda$ and $\omega$ must provide
consistent values for the single parameter $\varepsilon_{\text{pc}}$ or
$\alpha_{\text{m}}$. As noted in the previous section, the
STAR group analysis of the weight factor $\omega$
in the central Au+Au collisions at
$\sqrt{s_{NN}}=130$ GeV based on Model I indicates a value of
$\varepsilon_{\text{pc}}\simeq 0.8$, which corresponds to 
$\lambda\simeq 0.94$ \cite{STAR_3pi}. 
Such a large $\lambda$ seems to contradict the two-particle
correlation data. However, we make a correction for the long-lived
resonances in the two-pion correlation function before proceeding to
conclusive
calculations to obtain the parameters $\epsilon$ and $\alpha$ using
Eqs.~\eqref{eq:ome-pc}--\eqref{ome-pm}. Then we show that the
analysis of STAR is consistent with the value of $\epsilon^{\text{pc}}$
obtained from the resonance-corrected chaoticity.

\section{Long-lived resonance decay contributions to
 $\lambda$}\label{sec:lambda}

The existence of long-lived resonances that decay into pions causes the
apparent reduction of the chaoticity in the two-pion correlation
function. This is because the pion source caused by the long-lived
resonance decay spreads over a much larger scale than that which is
measurable with the
present experimental resolution.
 The reduction factor $\lambda^{\text{eff}}$ is related to the
fraction of pions from long-lived resonances as \cite{Csorgo_ZPHYS}
\begin{equation}
 \sqrt{\lambda^{\text{eff}}}
  =1-\frac{N_\pi^{\text{r}}}{N_\pi} \label{eq:lambdaeff},
\end{equation}
with $N_\pi$ being the total number of emitted pions
and $N_\pi^{\text{r}}$  being the number of pions emitted from long-lived
resonances.
We estimate the fraction of the $\pi^-$ originating from the
resonances with the help of a statistical model.
According to the recent analyses, this statistical model provides a
simple and good description of the particle yields of the RHIC
experiments \cite{Braun-Munziger_PLB518,Broniowski_PRL87}.
We take account of the resonances
up to $\Sigma^*(1385)$, in accordance with
Ref.~\cite{Wiedemann_PRC56}.  We treat 
$K_{\text{s}}^0, \eta, \eta', \phi, \Lambda, \Sigma$
 and $\Xi$ as the long-lived resonances whose widths are smaller than 5 MeV.
Heavier resonances can be ignored, because their contributions are
thermally suppressed.
The thermodynamic parameters are determined with a $\chi^2$ fitting to
the experimental data for the particle ratio. Here we take into account
$\bar{p}/p, \bar{p}/\pi^-, K^-/\pi^-, K^{*0}/h^-, \bar{K^{*0}}/K^{*0},
\bar{\Lambda}/\Lambda$ 
and $\bar{\Xi}/\Xi$. 
From the fit, we obtain the temperature
$T=158\pm 9$ MeV and the chemical potential with respect to the baryon number
$\mu_{\text{B}}=36 \pm 6$ MeV, with $\chi^2/{\text{dof}}=2.4/5$
\cite{Noteonmub}. For simplicity, we fix the chemical potential with
respect to the
third component of the isospin as $\mu_{I_3}=0$. This does not
affect the result, because the experimental data show $\pi^-/\pi^+ \simeq 1$.
The ratio of the number of particles $i$ to the number of particles $j$ can
be calculated using the number densities in the local rest frame
\cite{Cleymans_PRC60}.

Then, the true chaoticity $\lambda^{\text{true}}$, which is characterized
by the partial coherence of the source, is estimated as
$\lambda^{\text{true}}=\lambda^{\text{exp}}/\lambda^{\text{eff}}$, with
use of Eq.~\eqref{eq:lambdaeff}.
In general, $\lambda^{\text{exp}}$ obtained from the
1-dimensional Gaussian fitting can differ from that extracted from the
3-dimensional Gaussian fitting because of the projection,
experimental resolution, and other effects, though these should be the
same for an ideal measurement. In this paper, we use the value extracted
from the 3-dimensional Gaussian fitting 
\begin{equation}
 C_2^{\text{fit}}(\boldsymbol{q}) = 1 + \lambda^{\text{exp}}
  \exp\left( -R_{\text{side}}^2 q_{\text{side}}^2
      -R_{\text{out}}^2 q_{\text{out}}^2
      -R_{\text{long}}^2 q_{\text{long}}^2 \right),
  \label{eq:c2fit}
\end{equation}
as $\lambda^{\text{exp}}$.
The STAR collaboration measured \cite{STAR_HBT} $\lambda^{\text{exp}}$
for three transverse momentum bins. Because the transverse momentum
dependence of $\lambda^{\text{exp}}$ agrees with the results obtained from
Eq.~\eqref{eq:lambdaeff} \cite{Wiedemann_PRC56}, 
$\lambda^{\text{exp}}$ for the entire momentum range can be obtained by
computing an average of $\lambda^{\text{exp}}_i$ in which the square of
the number of pions is used as the weight:
\begin{equation}
  \overline{\lambda}^{\text{exp}} = \frac{\sum_{i=1}^{3} \lambda^{\text{exp}}_i
  \int_{i\text{-th bin}} k_t dk_t (dN/k_t dk_t)^2 }
  {\sum_{i=1}^{3} \int_{i\text{-th bin}} k_t dk_t (dN/k_t dk_t)^2}.
  \label{eq:lambdaav}
\end{equation}
Here, $\lambda^{\text{exp}}_i$ and $dN/k_tdk_t$ denote the measured
chaoticity in the $i$-th bin and the pion transverse momentum distributions,
respectively. We give a derivation of Eq.~\eqref{eq:lambdaav} in the
appendix. In the calculation of Eq.~\eqref{eq:lambdaav}, we use a
result of the Bose-Einstein fit to the pion spectra carried out by the STAR
collaboration \cite{STAR_Spectra}. 

Using Eq.~\eqref{eq:lambdaav}, we find
$\overline{\lambda}^{\text{exp}}=0.57\pm 0.06$. 

Combining $\lambda^{\text{eff}}$ obtained from
the statistical model calculation, we finally obtain
$\lambda^{\text{true}} = 0.93\pm 0.08$ for the present parameter region.
The uncertainty of 0.08 is the sum of the experimental uncertainty on
$\overline{\lambda}^{\text{exp}}$ calculated from uncertainties on each
$\lambda_i^{\text{exp}}$ and $dN/k_t dk_t$ and the statistical error
propagated from the fit of the thermodynamic quantities $T$ and
$\mu_{\text{B}}$ at 1-$\sigma$ confidence level.

\section{Combined analysis of the two- and three-particle correlation
 function}\label{sec:omega}
\begin{figure}[ht]
 \begin{center}
  \includegraphics[width=3.875in]{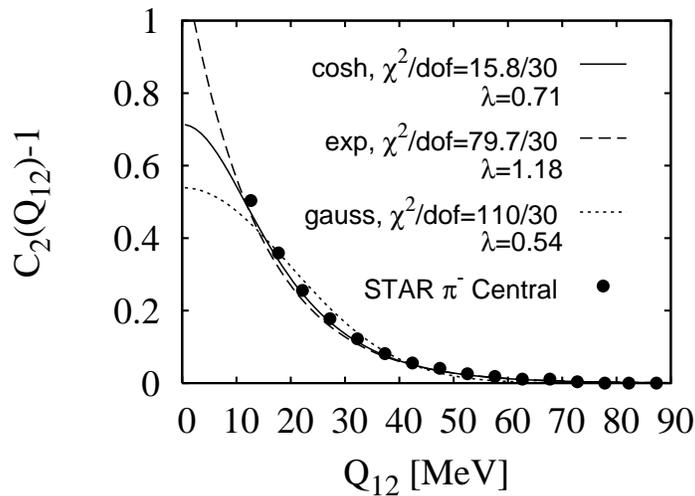}
  \caption{\label{fig:c2}Two-pion correlation function
  $C_2(Q_{12})$. The curves represent the results of the fits for
  each source function. The dots represent the
  experimental results taken from Ref.~\cite{STAR_HBT}.}
 \end{center}
\end{figure}

\begin{figure}[ht]
 \begin{center}
  \includegraphics[width=3.875in]{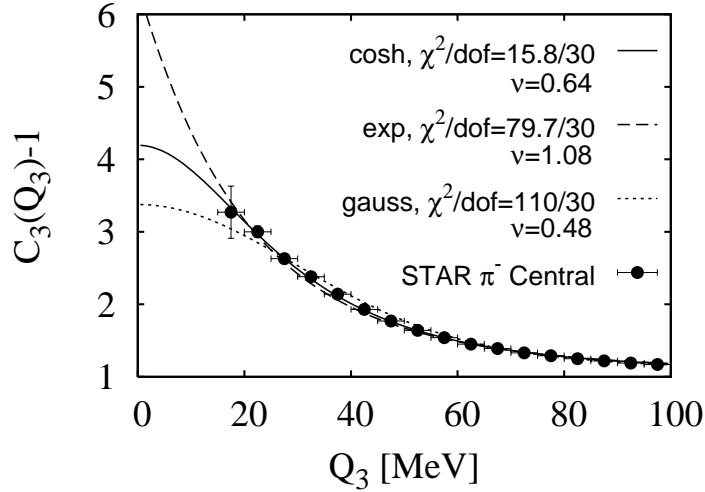}
  \caption{\label{fig:c3}Three-pion correlation function $C_3(Q_3)$. 
  Here, the curves and dots have the same identifications as in
  Fig.~\ref{fig:c2}. The experimental data are
  taken from Ref.~\cite{STAR_3pi}.}
 \end{center}
\end{figure}

In order to obtain the weight factor $\omega$, we have to
extrapolate $r_3(Q_3)$ [Eq.~\eqref{eq:r3}] to $Q_3=0$. 
Because $r_3(Q_3)$ is constructed from $C_{2}(Q_{ij})-1$ 
and $C_{3}(Q_3)-1$ [Eq.~\eqref{eq:r3}], we need to 
establish the functional forms of $C_2(Q_{ij})$ and 
$C_3(Q_3)$ so as to reproduce the experimental data
(Figs.~\ref{fig:c2} and \ref{fig:c3}). 
Although quadratic and quartic fits to $r_3(Q_3)$ at low $Q_3$ are used in
Ref.~\cite{STAR_3pi}, we adopt an alternative approach, because the
power law fits are relevant only for small $Q_3$, and it should be better
to extrapolate $r_3(Q_3)$ to $Q_3=0$ after the fits of $C_2$ and $C_3$
within finite but broad relative momentum ranges. We also note that
the $Q_3$ dependence of $r_3$ may be more complicated,
\cite{Nakamura_PRC60}
and the quadratic and quartic fitting 
tends to lead to an overestimated weight factor. 
In the following, we treat a set of $\pi^-$ data measured in the lowest
momentum bin for the event with the highest multiplicity.\cite{STAR_3pi} We
construct $C_2$ and $C_3$ from simple source functions with a set of
parameters. This construction is determined by minimizing $\chi^2$ with
respect to the
experimental data. For simplicity, we used a spherically symmetric
Fourier-transformed source function with simultaneous emission,
$F_{ij}=f_{ij}(|\boldsymbol{q}_{ij}|)e^{i(E_i-E_j)t_0}$,
where the exponential term corresponds to emissions
at a fixed time $t_0$. We believe that the assumption of instantaneous
emission is a good approximation, because recent two-particle
correlation data reveal a small time duration (\textit{i.e.}
$R_{\text{out}}\sim R_{\text{side}}$). In addition, the existence of a
finite time duration would not
affect the following treatment, because it does not influence the chaocitity
in the two-particle correlation function but, rather, decreases the
width of the
correlation function. For the
spatial part, $f_{ij}$, we consider a Gaussian, an
exponential, and a function written in terms of $\cosh$, which becomes
quadratic
at small $|\boldsymbol{q}|$ and exponential at large $|\boldsymbol{q}|$:
\begin{align}
 f_{1,ij}(|\boldsymbol{q}_{ij}|) &= 
 \exp\left(-R^2 |\boldsymbol{q}_{ij}|^2 /2 \right),\label{eq:fgauss} \\
 f_{2,ij}(|\boldsymbol{q}_{ij}|) &= 
 \exp\left(-R|\boldsymbol{q}_{ij}|/2 \right), \label{eq:fexp}\\
 f_{3,ij}(|\boldsymbol{q}_{ij}|) &= 
 \frac{1}{\sqrt{\cosh(R|\boldsymbol{q}_{ij}|)}}.\label{eq:fcosh}
\end{align}
Here, $\boldsymbol{q}_{ij}=\boldsymbol{p}_i-\boldsymbol{p}_j$, and $R$ is
the size parameter. The correlation functions are then calculated from
the relations
\begin{equation}
 C_2(\boldsymbol{p_1,p_2})=1+\lambda_{\text{inv}}\frac{f_{12}^2}{f_{11}f_{22}} 
  \label{eq:c2c}
\end{equation}
and
\begin{equation}
 C_3(\boldsymbol{p_1,p_2,p_3}) = 
 1+\nu\left(\sum_{(i,j)}\frac{f_{ij}^2}{f_{ii}f_{jj}}
 +2\nu_3 \frac{f_{12}f_{23}f_{31}}{f_{11}f_{22}f_{33}}\right) \label{eq:c3c},
\end{equation}
where $\lambda_{\text{inv}}$, $\nu$ and $\nu_3$ are phenomenological adjustable
parameters
accounting for non-trivial coherence effects,
and $\sum_{(i,j)}$ is a summation over $(i,j)=(1,2),(2,3),(3,1)$.
The forms Eqs.~\eqref{eq:c2c} and \eqref{eq:c3c} are chosen so as to
represent the chaotic case for $\lambda_{\text{inv}}=1$, $\nu=1$ and
$\nu_3=1$. Equation \eqref{eq:c2c} is a standard correlation function to
fit data \cite{STAR_HBT}. Although an alternative form of the
three-particle correlation function, $C_3=1+\nu \sum_{(i,j)}
\frac{f_{ij}^2}{f_{ii}f_{jj}} + 2\nu^{3/2}
\frac{f_{12}f_{23}f_{31}}{f_{11}f_{22}f_{33}}$,\footnote{Actually, the
resultant $\chi^2$ value for this expression is worse than that for
Eq.~{\eqref{eq:c3c}}.} is often used, this
equation corresponds to a perfectly chaotic source with a long-lived
resonance decay contribution, and is incapable of taking into account
non-trivial
coherence effects. Therefore, we simply parametrize the
three-particle correlation function by including the parameters from
both the 
two-particle correlation part [the second term in Eq.~\eqref{eq:c3c}] and
the genuine three-particle correlation part [the third term in
Eq.~\eqref{eq:c3c}]. Note that we need only $C_2(0)-1$ and $C_3(0)-1$
to obtain $\omega$. We can set $\nu_3=1$ without loss of
generality for a description of $C_3(Q_3)$ at small $Q_3$.
Then, the parameters $R$, $\lambda_{\text{inv}}$ and $\nu$ are
determined through a simultaneous $\chi^2$ fitting to both 
$C_2(Q_{12})$ and $C_3(Q_3)$. 

\begin{figure}[ht]
 \begin{center}
  \includegraphics[width=3.875in]{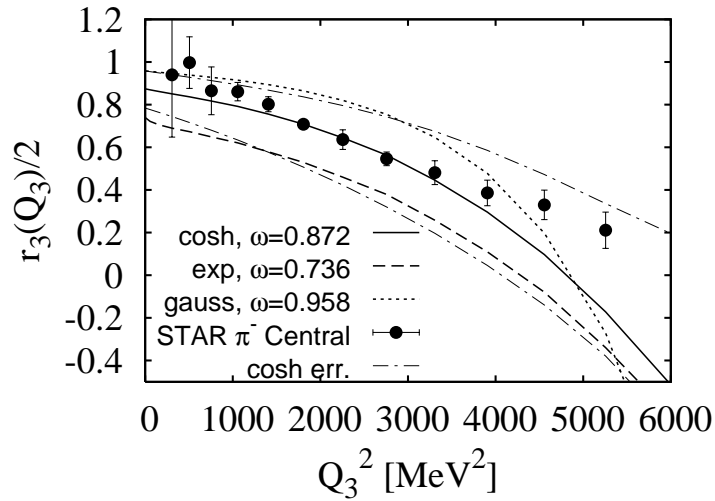}
  \caption{\label{fig:r3}The $Q_3$ dependence of $r_3 /2$.  The curves
  and dots have the same identifications as in Figs.~\ref{fig:c2} and
  \ref{fig:c3}, except for the dash-dotted curves,
  which represent the upper and lower bounds of the cosh
  case. The experimental data are taken from Ref.~\cite{STAR_3pi}. See
  the text for details.}
 \end{center}
\end{figure}

The three-particle correlator $r_3(Q_3)/2$ is then obtained from separate
projections of the numerator and denominator of Eq.~\eqref{eq:r3}
onto the single variable $Q_3$ \cite{WA98_3piprl}. 
The results are depicted in Figs.~\ref{fig:c2}--\ref{fig:r3} and
listed in Table \ref{tbl:fit}.

\begin{table}[ht!]
\caption{\label{tbl:fit}Results of the $\chi^2$ fitting to $C_2$ and $C_3$}
 \begin{ruledtabular}
 \begin{tabular}[t]{cccccc}
 $f(|\boldsymbol{q}|)$ & $R$ [fm] &  $\lambda$ &  $\nu $ &  
  $\chi^2 / \text{dof}$ & $\omega$\\  \hline

  $f_1 $& 7.0$\pm$0.07 & 0.54$\pm$0.01  & 0.48$\pm$0.01 & 110/30
  &0.958$\pm$0.09 \\

  $f_2$ & 14.4$\pm$0.2  &
  1.18$\pm$0.03 & 1.08$\pm$0.03 & 79.7/30 & 0.736$\pm$0.09 \\

  $f_3$ & 15.2$\pm$0.2 & 0.71$\pm$0.01 & 0.64$\pm$0.02 & 15.8/30 &
  0.872$\pm$0.097 \\
 \end{tabular}
 \end{ruledtabular}
\end{table}

The results show the failure of the Gaussian [Eq.~\eqref{eq:fgauss}] and 
exponential [Eq.~\eqref{eq:fexp}] fittings. In Figs \ref{fig:c2} and
\ref{fig:c3}, each curve looks in agreement with the data, but the values of $\chi^2/\text{dof}$ are
larger than in the $\cosh$ source function case. Furthermore, the
disagreement in $r_3(Q_3)$ for the exponential (Gaussian) case 
reveals an over-estimation (under-estimation) of $C_2$ in the
extrapolation to small $Q_{\text{inv}}$.
Constrastingly, the $\cosh$ source function [Eq.~\eqref{eq:fcosh}]
yields excellent agreement with the experimental data. Note that 
the resultant errorbars of $\pm 0.097$ for $\omega$ in the cosh case can
also be obtained by an extrapolation of the error band expressed by the
dash-dotted curves in Fig.~\ref{fig:r3}. The error band is calculated
using the propagation law from the errors of $R$, $\lambda_{\text{inv}}$
and $\nu$ in the $\chi^2$ fit (1-$\sigma$ confidence level).
Hence, our result for $r_3$ in the cosh case is consistent with all the
experimental data points, within the error bars.

The deviation
in Fig.~\ref{fig:r3} for larger $Q_3^2$ is not significant. Because both
the numerator and denominator in Eq.~\eqref{eq:r3} go to zero at the
region, the projection requires a careful treatment and is strongly affected
by unimportant tails. However, we do not
know the \textit{real} three-particle distribution which should be used
as a weight. Therefore we use Eq.~\eqref{eq:c3c} instead. We confirmed
that the 
most important quantity, $\omega=r_3(0)/2$, does not depend on the
choice of the weight.

\begin{table}[t]
 \caption{\label{tbl:result1}Results for Models I and II}
 \begin{ruledtabular}
  \begin{tabular}[t]{ccc}
   & Model I & Model II \\\hline
   From $\lambda$ & $\varepsilon_{\text{pc}} = 0.73\pm 0.14$ &
   $\alpha_{\text{m}}= 12.6\pm 14$ \\
   From $\omega$ & $ \varepsilon_{\text{pc}}= 0.65\pm 0.11$ & 
   $\alpha_{\text{m}} = 7.6\pm 9.7$
  \end{tabular}
 \end{ruledtabular}
\end{table}

Now let us evaluate how chaotic the pion sources are by applying the models
[Eqs.~\eqref{eq:ome-pc}--\eqref{ome-pm}] to the obtained values
$\lambda^{\text{true}}=0.93\pm0.08$ and $\omega=0.872\pm0.097$, 
where we adopt the
result of the $\cosh$ source function.
The allowed parameter regions of Models I and II are 
summarized in Table \ref{tbl:result1}.   
In these single-parameter models, $\epsilon_{\text{pc}}$ and
$\alpha_{\text{m}}$ should give consistent results for the two input
quantities, $\lambda^{\text{true}}$ and $\omega$. Table
\ref{tbl:result1} reveals that, after the appropriate correction for the
long-lived resonance decay contributions to the two-pion correlation
chaoticity index, the two-pion correlation function gives results consistent
 with the three-pion correlation function. The value of
$\epsilon_{\text{pc}}$ obtained from $\omega$ is slightly smaller than
that obtained in the
previous analysis carried out by the STAR collaboration \cite{STAR_3pi}, while
the values of 
$\epsilon_{\text{pc}}$ obtained from $\lambda^{\text{true}}$ and $\omega$ are
consistent with the result given in Ref.~\cite{STAR_3pi}. 
The very large uncertainty on $\alpha_{\text{m}}$
is due to rapid changes of $\alpha_{\text{m}}$ as a function of
$\lambda^{\text{true}}$ and $\omega$ in these parameter regions. This
feature may suggest the
existence of a chaotic background introduced in Models I and III.

The success of extracting the ``true'' chaoticity by considering the long-lived
resonance decay contribution suggests that it is needed to determine
the source of pions, i.e., direct and short-lived resonance
decay or long-lived resonance decay. Recently, a ``partial'' Coulomb
correction has been proposed \cite{Bowler_PLB270,Sinyukov_PLB432} and
actual computations have been carried out
\cite{CERES_NPA714,PHENIX_PRL93}, in which pions from long-lived
resonance decays are treated separately.
Such a correction affects not only the HBT
radii but also $\lambda^{\text{exp}}$. We leave its treatment for future
publications \cite{Morita_future}.

\begin{figure}[t]
 \includegraphics[width=5.875in]{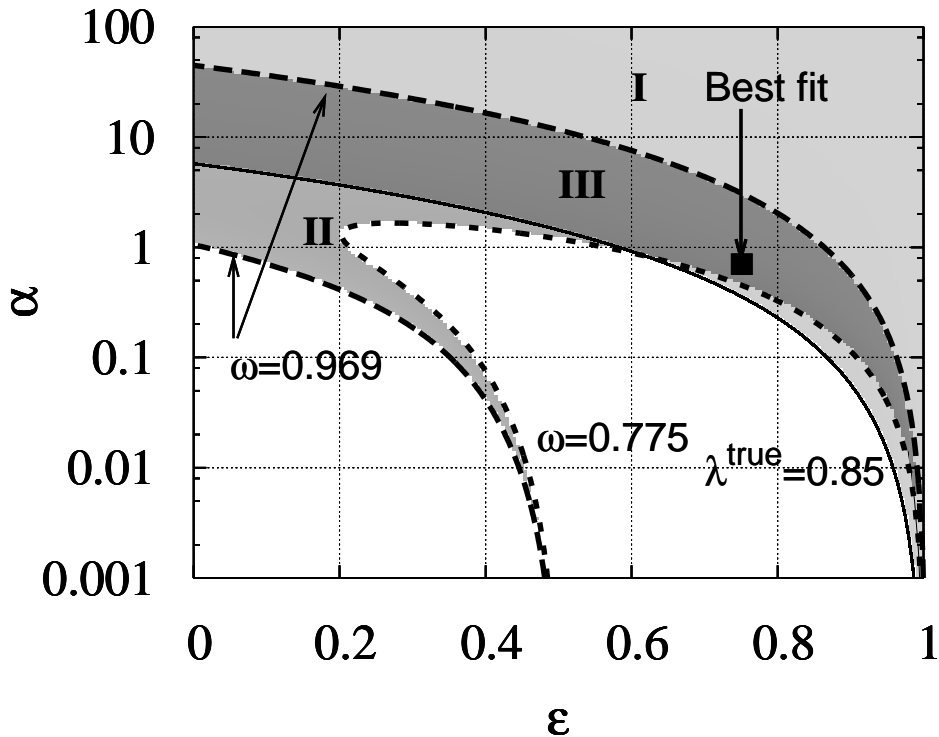}
 \begin{center}
  \caption{\label{fig:e-a}Results for Model III. Area I is the
  allowed parameter region corresponding to the value of
  $\lambda^{\text{true}}$ indicated by the solid line
  ($\lambda^{\text{true}}=0.85$). Area II, which is bounded by the
  dashed curve (upper bound of $\omega$) and dotted curve (lower bound
  of $\omega$), is
  the allowed parameter region corresponding to the range of $\omega$.
  Area III, which is characterized by overlap of areas I and II, is the
  allowed parameter region for $\alpha$ and $\varepsilon$. The filled
  solid represents the best fit point.}
 \end{center}
\end{figure}

The allowed region of $\varepsilon$ and $\alpha$ for Model III 
is displayed in Fig.~\ref{fig:e-a}. 
In this figure, the darkest shaded area, labeled ``III'', where the two
lighter areas overlap,
represents the ranges of $\varepsilon$ and $\alpha$ that correspond to the
regions of $\lambda_{\text{true}}$ and 
$\omega$ estimated from the experimental data.
The best fit point, $\varepsilon=0.75$ and $\alpha=0.77$, is indicated
by the solid square. Unfortunately, the uncertainty is still too large
to provide strong constraints on the pion production
mechanism. Nevertheless, the result shows a strong correlation between
$\alpha$ and $\varepsilon$; that is, the experimental data reveal a mostly
chaotic source which allows both a
``partially coherent'' picture ($\alpha\sim 0, \varepsilon \leq 1$) and
a ``multicoherent'' picture ($\alpha \gg 1, \varepsilon\sim 0$). 
Though a fully chaotic source, $\varepsilon =1$, is not excluded,
the strong $Q_3$ dependence of $r_3(Q_3)$ seen in Fig.~\ref{fig:r3} is
inconsistent with chaotic and symmetric sources. 
Even if the source is perfectly chaotic, the asymmetry of the source could
cause a strong $Q_3$ dependence. In this case, however, $r_3$ becomes
almost unity at small $Q_3$, while it deviates from unity at large $Q_3$
\cite{Nakamura_PRC60}. Hence, we can claim that small coherent
components are more likely to be produced.

In summary, we have given analyses of the two- and three-pion correlation
data obtained by STAR, in terms of three
kinds of partially coherent source models. We found that 
all three models give consistent description for both the
three- and two-pion correlations if we take account of the
apparent reduction of the chaoticity in the two-pion correlation
function caused by the long-lived resonance
decays. All models give the consistent picture 
that pion emission in the RHIC experiment is mostly chaotic, while
coherent components are likely to exist (see Table II and
Fig.~\ref{fig:e-a}). 
Further interesting problems concerning the
multiplicity and the collision energy dependence of the chaoticity 
will be published elsewhere \cite{Morita_future}.

\acknowledgments
 The authors would like to acknowledge Professor I.~Ohba and Professor
 H.~Nakazato for
 their support of this work.
 This work is partially supported by the
 Ministry of Education, Culture, Sports, Science and Technology,
 Japan (Grant No.~13135221)
 and Waseda University Grant for Special Research Projects (Nos.~2003A-095
 and 2003A-591).

\appendix

\section{Derivation of $\overline{\lambda}^{\text{exp}}$}

From Eq.~\eqref{eq:lambdaeff}, assuming that the $k_t$ dependence of
$\lambda^{\text{exp}}$ is governed by the long-lived resonance decay
contributions, we can write down $\lambda_i^{\text{exp}}$ for
midrapidity as 

\begin{equation}
 \lambda_i^{\text{exp}} 
  = 
  \frac{\int_{k^i_{t_{\text{min}}}}^{k^i_{t_{\text{max}}}} k_t dk_t 
  \left(\frac{dN_{\text{dir+short}}}{k_t dk_t}
  \right)^2}{\int_{k^i_{t_{\text{min}}}}^{k^i_{t_{\text{max}}}} k_t dk_t
  \left(\frac{dN_{\text{total}}}{k_t dk_t}\right)^2},
  \label{eq:lambdaexp1}
\end{equation}
where $N_{\text{dir+short}}$ and $N_{\text{total}}$ represent the sum of the
number of particles from direct emission and decay of short lived
resonances and the total number of emitted particles, respectively.
The values $k^i_{t_{\text{min}}}$ and $k^i_{t_{\text{max}}}$ are the
minimum and maximum for a given momentum window of the $i$-th
bin. Hence, $k^i_{t_{\text{min}}} = k^{i-1}_{t_{\text{max}}}$. Then, the
domain of
integration for the momentum acceptance,
$k_{t_{\text{min}}}\equiv k^1_{t_\text{min}} < k_t <
k^n_{t_{\text{max}}}\equiv k_{t_{\text{max}}}$ ,
with $n$ being the number of the momentum bin, can be divided as
\begin{equation}
 \int_{k_{t_{\text{min}}}}^{k_{t_{\text{max}}}} 
  = \sum_{i=1}^{n}\int_{k^i_{t_{\text{min}}}}^{k^i_{t_{\text{max}}}}.
  \label{eq:lambdaexp2}
\end{equation}
In terms of the above relation, $\overline{\lambda}^{\text{exp}}$
can be written as
\begin{equation}
 \overline{\lambda}^{\text{exp}} 
  = 
  \frac{\sum_{i=1}^{n}\int_{k^i_{t_{\text{min}}}}^{k^i_{t_{\text{max}}}}
  k_t dk_t 
  \left(\frac{dN_{\text{dir+short}}}{k_t dk_t}
  \right)^2}
  {\sum_{i=1}^{n}\int_{k^i_{t_{\text{min}}}}^{k^i_{t_{\text{max}}}} k_t dk_t
  \left(\frac{dN_{\text{total}}}{k_t dk_t}\right)^2}.
  \label{eq:lambdaexp3}
\end{equation}
Substituting Eq.~\eqref{eq:lambdaexp1} into Eq.~\eqref{eq:lambdaexp3}, we
obtain Eq.~\eqref{eq:lambdaav}.

\end{document}